\documentclass[conference]{IEEEtran}
\renewcommand{\baselinestretch}{0.87}
\IEEEoverridecommandlockouts

\usepackage{cite}
\usepackage{amsmath,amssymb,amsfonts} %Mathematical Symbol,Equation
\usepackage{algpseudocode}
\usepackage{algorithm}
\usepackage[inkscapelatex=false]{svg}
\usepackage{algpseudocode}
\usepackage{graphicx} %Adding picture and figure
\usepackage{float} %Figure Alignment
\usepackage{textcomp} %Text formatting
\usepackage{eso-pic}
\usepackage{relsize}
\usepackage{fancyhdr}
\usepackage{booktabs}
\usepackage{xcolor}
\usepackage{comment}
\usepackage{csquotes}
\def\BibTeX{{\rm B\kern-.05em{\sc i\kern-.025em b}\kern-.08em
T\kern-.1667em\lower.7ex\hbox{E}\kern-.125emX}}
    
\makeatletter
\def\ps@IEEEtitlepagestyle{%
  \def\@oddfoot{\mycopyrightnotice}%
  \def\@evenfoot{}%
}
\def\mycopyrightnotice{%
  \gdef\mycopyrightnotice{}% just in case
}

\usepackage{eso-pic}

\begin{document}
%\title{DistBlock-VANET: Distributed Blockchain-based Vehicular Ad Hoc Networks (VANETs) through SDN-NFV for Smart City}

%\title{DistB-VANET: Distributed Blockchain-based Vehicular Ad-Hoc Networks through SDN-NFV for Smart City}

\title{DistB-VNET: Distributed Cluster-based Blockchain Vehicular Ad-Hoc Networks through SDN-NFV for Smart City}

%{\footnotesize \textsuperscript{}}
%\thanks{}
%}

\begin{comment}
\author{\IEEEauthorblockN{Anichur Rahman \IEEEauthorrefmark{1}, Md. Jahidul Islam\IEEEauthorrefmark{2}, Sumaiya Kabir\IEEEauthorrefmark{3}, and Mostofa Kamal Nasir\IEEEauthorrefmark{4}}
\IEEEauthorblockA{\textit{Dept. of Computer Science and Engineering} \\
\textit{Green University of Bangladesh, Mawlana Bhashani Science and Technology University} \\
jahidul.jnucse@gmail.com\IEEEauthorrefmark{1}, 
anis.mbstu.cse@gmail.com\IEEEauthorrefmark{2},
sumaiya@cse.green.edu.bd\IEEEauthorrefmark{3} and kamal@mbstu.ac.bd\IEEEauthorrefmark{4}}}
\end{comment}

\author{\IEEEauthorblockN{Anichur Rahman\IEEEauthorrefmark{1}, MD. Zunead Abedin Eidmum\IEEEauthorrefmark{2}, Dipanjali Kundu\IEEEauthorrefmark{3}, Mahir Hossain\IEEEauthorrefmark{4},\\
MD Tanjum An Tashrif\IEEEauthorrefmark{4}, Md Ahsan Karim\IEEEauthorrefmark{5}, and Md. Jahidul Islam\IEEEauthorrefmark{6}}

\IEEEauthorblockA{
\textit{Dept. of Computer Science and Engineering, National Institute of Textile Engineering and Research (NITER)},\\
\textit{Constituent Institute of Dhaka University, Savar, Dhaka-1350}\\
\textit{Dept. of Internet of Things and Robotics Engineering, Bangabandhu Sheikh Mujibur Rahman Digital University},\\
\textit{Dept. of Computer Science and Engineering, Green University of Bangladesh}\\
anis.mbstu.cse@gmail.com\IEEEauthorrefmark{1},
1801031@iot.bdu.ac.bd\IEEEauthorrefmark{2},
dkundu@niter.edu.bd\IEEEauthorrefmark{3},
mahihossain114@gmail.com\IEEEauthorrefmark{4},\\
mtashrif20@niter.edu.bd\IEEEauthorrefmark{5},
makarim11@niter.edu.bd\IEEEauthorrefmark{6},
jahid@cse.green.edu.bd\IEEEauthorrefmark{7}}
}

\maketitle
%\conf{International Conference on Innovation in Engineering and Technology (ICIET) 23-24 December 2019}
\maketitle

%%%%%%%%%%%%%%%%%%%%%%%%%%%%%%%%%%%%%%%%%%%%%%%%%%%%%%%%%%%%%%%%%%%%%%%%%%%%%%%%
\begin{abstract}
\boldmath
\textcolor{black}{In the developing topic of smart cities, Vehicular Ad-Hoc Networks (VANETs) are crucial for providing successful interaction between vehicles and infrastructure. This research proposes a distributed Blockchain-based Vehicular Ad-hoc Network (\enquote{DistB-VNET}) architecture that includes binary malicious traffic classification, Software Defined Networking (SDN), and Network Function Virtualization (NFV) to ensure safe, scalable, and reliable vehicular networks in smart cities. The suggested framework is the decentralized blockchain for safe data management and SDN-NFV for dynamic network management and resource efficiency and a noble isolation forest algorithm works as an IDS (Intrusion Detection System). Further, \enquote{DistB-VNET} offers a dual-layer blockchain system, where a distributed blockchain provides safe communication between vehicles, while a centralized blockchain in the cloud is in charge of data verification and storage. This improves security, scalability, and adaptability, ensuring better traffic management, data security, and privacy in VANETs. Furthermore, the unsupervised isolation forest model achieves a high accuracy of 99.23\% for detecting malicious traffic. Additionally, reveals that our method greatly improves network performance, offering decreased latency, increased security, and reduced congestion, an effective alternative for existing smart city infrastructures.}

\end{abstract}
\vspace{2mm}

%\IEEEpeerreviewmaketitle

\section{Introduction}
In recent years, the field of SDN and NFV has made significant advancements in the context of smart city systems. Previous studies \cite{ali2023reliable, rahman2024blocksd, rahman2024internet} have explored this field of technology. With the increasing population, wireless connectivity has advanced, and research in VANETs has expanded, a critical component of smart city wireless systems. VANETs are designed to enable communication between vehicles and their operators to improve traffic safety and efficiency. VANET, a sub-category of the mobile ad hoc network, serves the function of the transmission of data between the vehicle and the vehicle operator.%, as shown in Fig. \ref{fig:f1}.}

\begin{comment}
\begin{figure}[h]
    \centering
    \includegraphics[scale=0.4]{Figures/Intro.jpg}
   \caption{Hierarchy of VANET \cite{dak2012literature}}
    \label{fig:f1}
\end{figure}
\end{comment}

\vspace{2mm}
In contrast to traditional VANET technology, there is an excellent opportunity for Blockchain that provides better security for the vehicle to the vehicle transmission system. As in the conventional system, the data from the devices are stored in a centralized server, and from this server, all other nodes collect and use the information required \cite{rahman2023towards}. This centralized server may be the main attraction for hackers. Because the information stored here is very critical for efficient communication between the nodes of the transport system, this information requires a secure platform where the data collected is stored securely and the data preserved effectively. Again, there is another emerging technology, Blockchain, which is a distributed decentralized technology, and the information is passed as a block. Here, only authorized block can take their place in the chain, thus maintaining the privacy of the blocks of information. The idea of a decentralized blockchain may, therefore, be a potential solution for VANET technology \cite{mendiboure2020survey}. To effectively transfer information from one vehicle to another by storing it in a trusted location, leveraging the design of the Blockchain distributed process \cite{fang2024decentralized}.

\vspace{2mm}
There are many critical issues that are still unsolved in VANETs, mainly in security, privacy, and time management. Communication delays between vehicles could cause accidents and hinder traffic management. Additionally, challenges like the high power demands for node creation and block authorization affect vehicular management systems \cite{hakiri2024joint}. Since cities are evolving in a smart way, efficient traffic management has become necessary, so improved accuracy, trust, transparency, and security are required. Technologies like SDN, NFV, AI, and Big Data can enhance performance, due to SDN offering a layered platform to manage increasing data from vehicle-to-vehicle communication \cite{rahman2021smartblock}. However, centralized VANET architectures remain vulnerable due to centralized data storage, making them vulnerable to hacking. Delays or system failures could seriously impact traffic control. Despite efforts to improve them, a fully secure and reliable solution is still needed.

\vspace{2mm}
\textcolor{black}{As smart cities evolve, there is a growing need for advanced technologies that can handle complex urban systems more efficiently. In this context, efficient traffic management plays a pivotal role in ensuring safety and reducing congestion. Our proposed system combines SDN, NFV, and Blockchain to provide superior accuracy, trust, transparency, latency, and security. SDN, with its layered architecture, offers an effective platform to handle large volumes of data generated by vehicular networks. However, the reliance on centralized servers in SDN makes the system vulnerable to attacks. Integrating NFV helps mitigate these risks by decentralizing traffic management, reducing power consumption, and enhancing flexibility. In this system, NFV ensures that resources are allocated more efficiently, improving overall performance \cite{rahman2022sdn}.
The combination of SDN, NFV, and Blockchain could solve the conventional problems in VANET. However, there are still some research gaps about how to successfully integrate these technologies to find a better solution to the open challenges of the smart vehicle management system. Before incorporating SDN, the security challenges of SDN need to be mitigated. Some of the attacks of SDN, namely Denial of Service attacks, flooding attacks, and some others addressed by researchers \cite{arif2019survey, bhatia2019software}. At the same time, NFV is rather not free from technical challenges \cite{bradai2020software}. Monitoring the flow of information, failure recovery, especially in the telecommunication system, and allocation of available resources are some of the significant issues in NFV-related technologies \cite{chowdhury2019re}}. \vspace{2mm}

\textcolor{black}{This research introduces a novel approach to enhance traditional VANET systems by integrating Blockchain with SDN and NFV in smart city environments. Blockchain’s decentralized structure ensures secure vehicular data transmission while reducing the risk of central points of failure. In this system, clusters of vehicles are formed, with a cluster head communicating with distributed nodes linked to a central cloud. While Blockchain enhances security, challenges like high computational demands and delays need addressing. By combining SDN and NFV, this approach optimizes both security and performance in smart city VANETs. This work's contributions are as follows:\vspace{1mm}}
 
\begin{itemize}

    \item The authors propose a distributed Blockchain-based Vehicular Ad-hoc Network \enquote{DistB-VNET} architecture that provides security, scalability,  reliability and confidentiality as well as better traffic management in the VANET environment .\vspace{1mm}
    
    %\item \textbf{\enquote{DistB-VNET}} scheme has been proposed by the authors, which provides security, scalability, and reliability as well as better traffic management in Vehicular Ad-Hoc Networks (VANET).
    %\item \textcolor{black}{We employ an unsupervised model isolation forest after the cluster head to detect and block malicious traffic at the edge level.}\vspace{1mm}

     \item An SDN technique has been integrated with multiple controllers from the data layer to the application layer to divide the load equally between the devices and controllers. Also, NFV is employed to ensure the automatic allocation of network resources efficiently.\vspace{1mm}
    
    \item \textcolor{black}{Additionally, an unsupervised model Isolation Forest is employed by the authors after the cluster head to detect and block malicious traffic in the edge level.}\vspace{1mm}
   
\end{itemize}
\vspace{2mm} 

\textbf{Organization:} The rest of the paper has been formed as follows: We studied and discussed the related literature in section II. Then, section III performs the proposed model for smart cities. The result analysis and discussion are assessed in section IV. In addition, section V presents the conclusion with future directions.

\section{Related Works}
Recently, researchers have addressed enormous work based on emerging leading technologies such as SDN, distributed Blockchain, NFV, and VANET technologies. In this section, the authors present an overview of some recent works in literature. \vspace{2mm}

Hemani et al.\cite{hemani2024designing} proposed a tamper-proof and transparent data sharing system between the nodes of autonomous  vehicles. The privacy of the shared information was preserved using smart contracts with solidity. Zalte et al.\cite{zalte2022synergizing} suggested a integrating blockchain technology to solve problems with data dispersion in VANETs. In addition, this research covered a blockchain that uses AI and data analytics. Diallo et al. \cite{diallo2022scalable} proposed a new system for managing traffic-related data in VANETs (Vehicular Ad hoc NETworks) using blockchain technology in their work. In their study, Abdullah et al. \cite{alharthi2021privacy} proposed a framework called Biometrics Blockchain (BBC) to secure data sharing between vehicles in Vehicular Ad-hoc Networks (VANETs). The BBC framework used biometric information to verify the identity of vehicles without revealing their actual identity. Hailin et al.  \cite{9875052} explored the integration of blockchain technology with Digital Twins (DTs) to map real-world traffic scenarios into a virtual space in VANETs to enhance intelligent transportation and employ blockchain for secure data storage and transmission in smart cities. The proposed model demonstrated lower average delay times, stable data message delivery rates, and high network security.\vspace{2mm}

Additionally, with this traditional VANET technology, other emerging technologies like SDN, Blockchain, NFV, Artificial Intelligence, Cloud computing, and so many other available technologies were incorporated together to solve and enhance the transportation management performance system. In their research, SULEYMAN et al.\cite{turner2023promising} investigated how Blockchain (BC) and SDN operate together in the context of the IoT \cite{rahman2023icn}. The six main implementation objectives of security, computing paradigms (fog and edge), trust management, access control \& authentication, privacy, and networking were used to categorize pertinent studies. Tahani et al.\cite{gazdar2022decentralized} in their article proposed a decentralized Blockchain-based trust management framework (BC-TMF) to compute trust metrics for vehicles and achieved better accuracy for malicious vehicles. In their study, Balaji et al.\cite{balaji2023networking} discussed the difficulties with traffic management and energy efficiency in VANETs. In addition, the authors presented the SDNTFP-C method, which combined Deep Learning (DL) models with the scalability, flexibility, and adaptability of SDN controllers.

\begin{table}[!htb]
\centering
\scriptsize
\tiny
\caption{Comparison of State-of-Arts}
\begin{tabular}{|p{1.5cm}|p{3.0cm}|p{3.0cm}|}
\hline
\textbf{Works} & \textbf{Contributions} & \textbf{Research Gap} \\ \hline
Hemani et al.(2024)\cite{hemani2024designing} & Implemented a smart contract with solidity to reserve information privacy for AV & Excessive power consumption  \\ \hline
Zalte et al. (2022)\cite{zalte2022synergizing} & Solved data dispersion problems in VANETs. & Computational power and Storage capacity Limitation \\ \hline
SULEYMAN et al. (2023)\cite{turner2023promising} &  Combined Blockchain and SDN together in the Context of IoT &Lack of regulatory frameworks and greater complexity. \\ \hline
Tahani et al. (2022)\cite{gazdar2022decentralized} & Computed trust metrics for vehicles and  acheived better accuracy& Privacy of vehicle owners is not protected \\ \hline
Balaji et al. (2023) \cite{balaji2023networking}  & Integrated deep learning (DL) models with the scalability, flexibility, and adaptability of SDN controllers & Lack of robustness\\ \hline
\end{tabular}

\label{tab:comparison}
\end{table}

\vspace{2mm}
In summary, the current research is not sufficient, and there are still many challenges in this system. Table \ref{tab:comparison} shows the contribution and the limitations of the previous studies, which we are going to solve.

\section{Proposed \enquote{DistBlock-VNET} Architecture for Smart Cities}

\begin{figure*}
\centering
\includegraphics[scale=0.85]{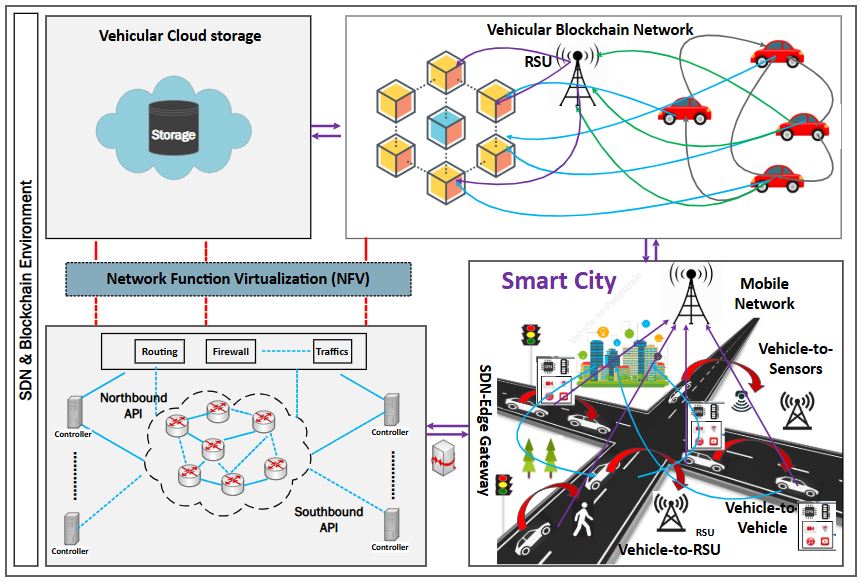}
 \caption{Proposed \enquote{DistB-VNET} Architecture for Smart City}  
 \label{fig:SDNIoTArc}
\end{figure*}

This section provides the architecture and operational framework of \enquote{DistB-VNET}, a distributed blockchain-based VANET combined with SDN and NFV  for smart cities. Fig. \ref{fig:SDNIoTArc} shows the integration of vehicular cloud storage, blockchain, and SDN controllers. Vehicular cloud storage remains as a centralized repository for traffic data, vehicle IDs, and sensor information, enhancing data sharing among vehicles and city services. The blockchain network functions as a decentralized ledger, providing data integrity by storing transactions immutably. SDN offers a programmable network architecture, dividing control and data planes for centralized network management. SDN controllers regulate data flow, routing, and quality of service using northbound and southbound APIs. Vehicles collect data by the sensors, upload it to vehicular cloud storage for processing, and then send it to the blockchain for validation. Network services, managed by NFV, handle the validated data, while SDN controllers take real-time decisions on routing, traffic management, and resource allocation. This integration provides a secure, scalable, and efficient vehicular communication network for smart cities context.

\subsection{Detection and Block Malicious Traffic}
\textcolor{black}{In this architecture, IoT devices send traffic to the distributed nodes through the cluster heads. However, it may be vulnerable to cyber-attacks executed in the form of network traffic. Traditional intrusion detection systems or supervised models can only detect known types of cyber-attacks like DDoS, SQL injection, etc. Moreover, in real-world scenarios, the attacks may be new, as methods of cyber-attacks evolve daily. To address this issue, instead of training a supervised model, we train an unsupervised model. This unsupervised model, named Isolation Forest, is trained on the IDS 2018 Intrusion dataset \cite{IDS2018_Kaggle}. The trained model functions as a primary verification system between IoT edge devices and distributed nodes. The Isolation Forest model checks all traffic, categorizing it as either benign or malicious. Malicious traffic is blocked by the verification process, and it is stored in temporary storage to prevent it from passing until it is cleared of malware \cite{rahman2023impacts}.
}

\subsection{Architectural Design for VANET with SDN-NFV Procedure}
In the context of smart cities, SDN emerges as a critical technology by separating network control from data forwarding, thereby permitting dynamic traffic management. By using SDN’s capabilities, traffic systems may adjust in real-time, enhancing efficiency and responsiveness. Integrating SDN with NFV further boosts system flexibility and scalability, enabling optimal resource allocation and minimizing operational expenses. The proposed architecture shown in Fig. \ref{fig:SDNIoTArc} corresponds with SDN and NFV standards, supporting dynamic network management through components such as the SDN Network Slice Manager (NSM), Network Slice Orchestrator (NSO), and Vehicular Application Manager (VAM). In vehicle networking situations like VANET and the Internet of Vehicles (IoV), SDN, NFV, and Fog Computing play significant roles in facilitating connection and optimal resource usage. Network slicing provides for specialized networks with specified Quality of Service (QoS) needs, while protocol selection, such as AODV, DSDV, or OLSR, depends on network characteristics and mobility patterns. The integration of SDN and NFV offers benefits such as flexibility, scalability, and cost reduction, virtualizing network services to enable for effective resource usage and dynamic allocation based on traffic conditions. These components provide dynamic network management and orchestration, providing optimal network performance and dependable vehicular communication \cite{rahman2020distb}.

\begin{figure*}[!ht]
    \centering
    \includegraphics[scale=0.25]{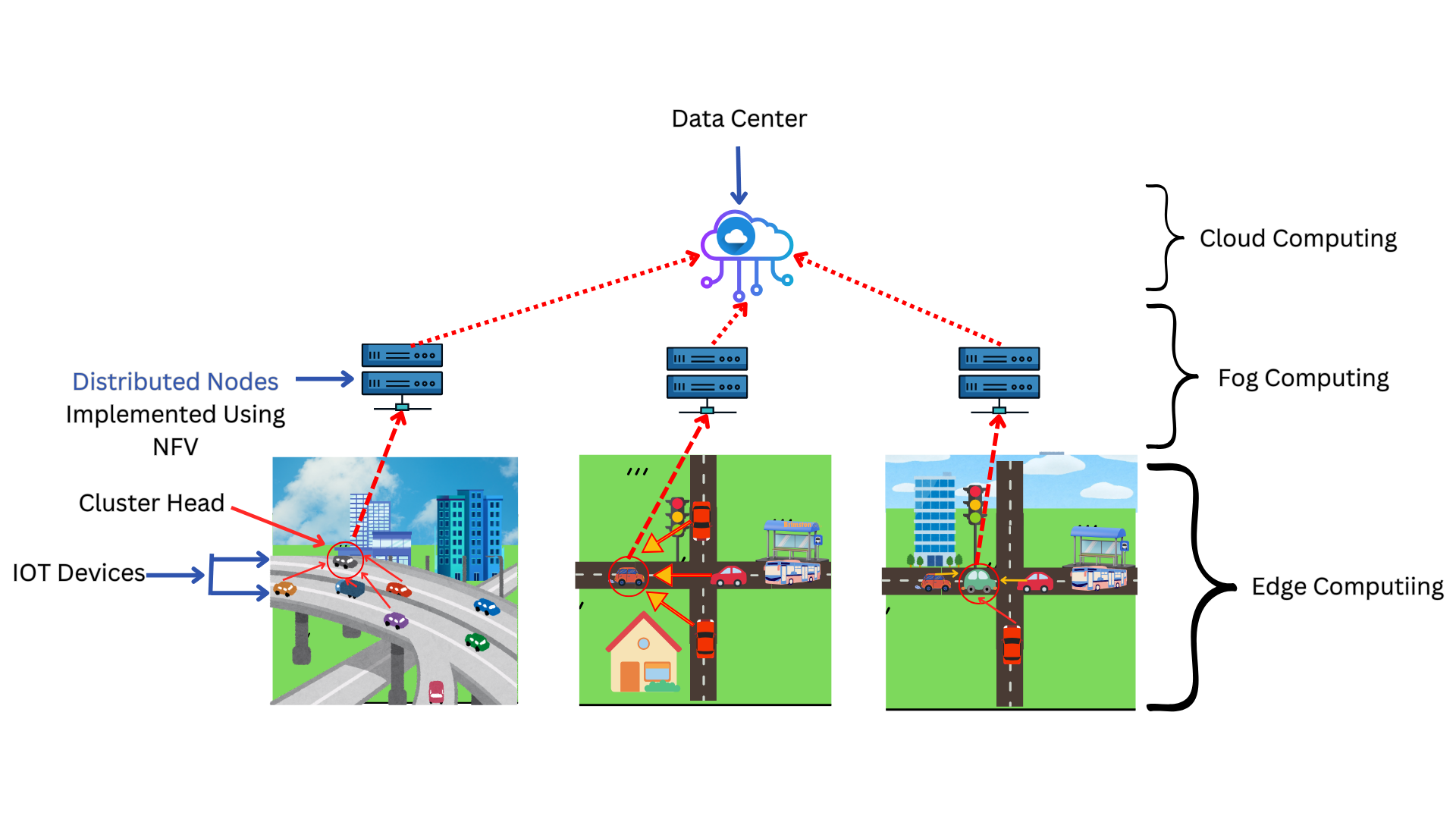}
    \caption{Data Collection Scenario from the Edge Layers}
    \label{fig:SDNIoTNFV}
\end{figure*}

\subsection{Proposed Algorithm for \enquote{DistB-VNET}}
\textcolor{black}{In this section, Alg. \ref{alg:distb-vnet} is presented for our proposed Distributed Blockchain-based Vehicular Ad-hoc Network (DistB-VNET), integrated with SDN and NFV technologies for smart city environments. The algorithm is designed to handle secure communication, vehicle clustering, and efficient resource management in the network. The \enquote{DistB-VNET} system first measures signal strengths and processing power to make clusters via Roadside Units (RSUs) where cluster heads lead communications. Messages are encrypted using the public-private key pair and verified through blockchain to ensure integrity. They have finally been forwarded by the SDN controllers. Additionally, the controller manages traffic and allocates resources automatically using NFV to optimize performance and reduce latency.}

\begin{algorithm}[!htb]
\caption{DistB-VNET: Secure Communication and Clustering Algorithm}
\scriptsize
\label{alg:distb-vnet}
\begin{algorithmic}[1]
    \State \textbf{Initialization:}  
    \Statex \hspace{1.5em} $V$: Set of vehicles 
    \Statex \hspace{1.5em} $C$: Set of clusters 
    \Statex \hspace{1.5em} $R$: Set of Roadside Units (RSUs) 
    \Statex \hspace{1.5em} $S$: SDN controller 
    \Statex \hspace{1.5em} $K_{pub}(v)$: Public Key of vehicle $v$
    \Statex \hspace{1.5em} $K_{priv}(v)$: Private Key of vehicle $v$
    \Statex \hspace{1.5em} $B$: Blockchain ledger
    \Procedure{Cluster Formation}{$V, R$}
        \For{each $v \in V$}
            \State $S(v), P(v)$ \Comment{Measure signal strength and processing power}
            \State $v \to r \in R$ \Comment{Broadcast to nearest RSU}
            \State $CH \gets \text{Elect}(R)$
            \State $C \gets \text{Create Cluster}(CH)$
            \State $(K_{C_{pub}}, K_{C_{priv}}) \gets \text{KeyPair}(C)$
        \EndFor
    \EndProcedure

    \Procedure{Secure Communication}{$CH, S, B$}
        \For{each $m$ from $v$ to $CH$}
            \State $H(m) \gets \text{Hash}(m)$
            \State $E(m, K_{priv}(v))$ \Comment{Encrypt with $K_{priv}$}
            \State $m \gets D(E(m), K_{pub}(v))$ \Comment{Decrypt with $K_{pub}$}
            \State $H(m), K_{C_{priv}} \to S$
            \State $\text{Validate}(S, B)$
        \EndFor
        \State $CH \to S$ \Comment{Forward data}
    \EndProcedure

    \Procedure{Resource Allocation}{$S, NFV$}
        \State $S \gets \text{Collect State Info and Traffic Load}$
        \State $\text{Allocate}(NFV)$ \Comment{Optimize latency and performance}
        \State $\text{Manage Traffic}(S)$
    \EndProcedure

\end{algorithmic}
\end{algorithm}

\subsection{Vehicles Controlling via Distributed Blockchain and SDN}
The combination of blockchain, SDN-VANET, IPFS and clustering provides a secure and scalable solution for smart city traffic management. Roadside Units (RSUs) transmit cluster data through a blockchain-based voting system. That safely selects the cluster head according to the vehicle type. As seen in Fig \ref{fig:SDNIoTNFV}, the cluster head then secures the communication by creating cryptographic key pairs and signing messages, which are subsequently confirmed by the SDN controller via blockchain. By assembling vehicle requests through cluster heads, who then hash and forward messages to the SDN controller, clustering lowers networking load \cite{rahman2020block}. The controller ensures a two-step verification process and recives the original messages and contents from IPFS. Blockchain ensures communication integrity within clusters by using public key infrastructure for vehicle data verification. Blockchain guarantees tamper resistance and transparency by securing the storage of cluster keys, traffic data, and vehicle data. SDN provides efficiant routing, resource management and centralized network control. IPFS facilitates distributed storage that guarantees scalability, While clustering improves communication by reducing latency and congestion. Interaction between RSU cluster members and SDN controllers is secured by cryptographic key management and secure message authentication. Guarantee the reliability and integrity of the system.
\subsection{Real Life Application of \enquote{DistB-VNET} Framework in Smart Cities}
Smart cities are regulated around huge volume of data generated by smart vehicles, IoT devices, sensors, actuators etc. As the concept of smart cities become popular, the issues of data safety, security becomes top priority. Distributed nature of blockchain ensures that all the data available is stored in multiple places so that they are less vulnerable to cyber attacks. For vehicular communication, it is important to preserve the privacy of the vehicle owner while sharing vehicle information. This problem can be also solved by using the smart contracts. To avoid collision and fast and reliable data transfer, the latency should be minimal and throughput should be maximum. \enquote{DistB-VNET} framework ensures the minimal latency by using NFV automatic resource allocation process. Finally, this framework ensures data privacy, safety and secure communication of information for smart vehicles which make it very much suitable for real life application in the smart cities.

\section{Result Analysis and Design}
{VANET is an important part of the smart city. Implementing blockchain to secure the network connection between smart vehicles in smart cities is a new addition, where the networking process can be taken to the next level using the combination of SDN-NFV with Edge and Fog computing. The persisting previous work of researchers in the field of Smart cities, Ad-Hoc networks, and blockchain inspired us to do this work. We used two types of blockchain technology to secure our network, distributed and centralized. Our smart vehicle connects with the fog network which later connects with the cloud server. The SDN technology is used to control the network part and NFV is used to virtualize the process. The result of our model is promising and excellent. This result means that our proposed model can be used in modern smart cities.}

\begin{table*}[!htb]
    \centering
    \scriptsize
    %\tiny
    \caption{Comparison of Proposed and IEAOCGO-C \cite{elhoseny2023intelligent} methods across various metrics}
    \label{tab:net_comparison}
    \renewcommand{\arraystretch}{1.2}
    \begin{tabular}{|c|c|c|c|c|c|c|c|c|c|c|}
        \hline
        \textbf{Vehicles} & \multicolumn{2}{c|}{\textbf{NLT (rounds)}} & \multicolumn{2}{c|}{\textbf{PDR (\%)}} & \multicolumn{2}{c|}{\textbf{THRPT (kbps)}} & \multicolumn{2}{c|}{\textbf{ETED (mJ)}} & \multicolumn{2}{c|}{\textbf{ECM (ms)}} \\
        \cline{2-11}
         & \textbf{Proposed} & \textbf{IEAOCGO-C} & \textbf{Proposed} & \textbf{IEAOCGO-C} & \textbf{Proposed} & \textbf{IEAOCGO-C} & \textbf{Proposed} & \textbf{IEAOCGO-C} & \textbf{Proposed} & \textbf{IEAOCGO-C} \\
        \hline
        20 & 5200 & 4500 & 99.37 & 94.14 & 71.23 & 63.25 & 6.04 & 8.14 & 30.96 & 34.22 \\
        \hline
        30 & 4900 & 4400 & 89.25 & 86.34 & 78.68 & 69.43 & 6.06 & 9.04 & 51.68 & 62.32 \\
        \hline
        40 & 4700 & 3900 & 86.45 & 85.13 & 83.14 & 74.45 & 6.09 & 9.45 & 70.19 & 82.33 \\
        \hline
        50 & 4200& 3600& 84.32& 83.42&89.23 & 79.66&7.45 &10.32 & 85.41& 105.46\\
        \hline
    \end{tabular}
    
\end{table*}

\subsection{Measurement Parameters}
\textcolor{black}{We have used various parameters such as gas consumption rate, throughput based on a number of vehicles, End- to-End Delay, Packet Delivery Ratio, Overhead,  and throughput based on cluster size, to measure the performance of the proposed framework.}

Throughput in vehicular network, considering the cluster size, is given by following equation where  $T(C)$ is the throughput (in Mbps) based on cluster size, $C$ is the size of the cluster, $D(C)$ is the average data generated by each vehicle in the cluster, $t_b$ represents the delay due to blockchain operations, $t_s$ is the time delay from SDN operations, $t_n$ accounts for NFV-related delays such as resource allocation, $t_c$ refers to the intra-cluster communication delay.

\begin{equation}
T(C) = \frac{C \times D(C)}{t_b + t_s + t_n + t_c}
\end{equation}

Moreover, throughput in a blockchain considering the number of vehicles, is given bellow where $T(V)$ is the throughput (in Mbps) based on the number of vehicles where $V$ represents the total number of vehicles in the network, $D(V)$ is the average data generated per vehicle, $t_b$ refers to the delay due to blockchain operations, such as transaction validation and block propagation, $t_s$ is the time delay from SDN operations, including routing and traffic management, $t_n$ accounts for the time delay due to NFV operations, $t_v$ is the communication delay between vehicles and infrastructure..  
\begin{equation}
T(V) = \frac{V \times D(V)}{t_b + t_s + t_n + t_v}
\end{equation}
The gas consumption rate \( G(T_x) \) depends on the number of transactions \( T_x \) and can be expressed as:
\begin{equation}
G(T_x) = G_0 \times T_x + C_b \times T_x
\end{equation}
Where \(T_x\) denotes the total number of transactions processed in the blockchain, \( G_0 \) is the base gas cost required for validating a single transaction and \( C_b \) represents the additional gas consumption due to the complexity of blockchain operations.

\subsection{Environment Setup}
\textcolor{black}{We have used a network emulator named Mininet and Ethereum for simulating our proposed methodology \cite{islam2021blockchain}. The parameters used for this simulation are shown in Table \ref{fig:SE}.} 

\begin{table}[h]
\caption{\textcolor{black}{Simulation Environment}}
 \label{fig:SE}
\centering
\scriptsize
\begin{tabular}{ p{3.6cm}|p{3.6cm} }
\hline

\textbf{Parameter} & \textbf{Values}\\
\hline
Blockchain Platform & Ethereum \\
Emulator & Mininet(version : 2.2.1)\\
Platform Type & Decentralized \\
Environment & Truffle, Ganache \\
Language & Solidity\\
Number of Nodes & 80 \\
Max. of deceleration & 5\( m/s^2 \) \\
Max. Acceleration & 3.6 \( m/s^2 \)\\
Max. vehicle  speed& 50\( m/s^2 \)  \\
Number of RSU Unit & 10 \\
Packet Size & 100-512 bytes \\
Number of transactions & Variable\\
Block size & Variable \\
Transaction per Block & Variable \\
\hline
\end{tabular}
\end{table}

\subsection{\textcolor{black}{Performance Evaluation}}
\textcolor{black}{Table \ref{tab:isolation forest} indicates the performance scores for our isolation forest algorithm. It achieves high accuracy of 99.23\% following by high precision of 99.14\%, recall of 99.15\% anf f-score of 99.07\%.}
\begin{table}[ht]
\centering
\scriptsize

\caption{Performance Metrics of Isolation Forest}
\begin{tabular}{|c|c|c|c|}
\hline
\textbf{Accuracy (\%)} & \textbf{Precision (\%)} & \textbf{Recall (\%)} & \textbf{F1-Score (\%)} \\ \hline
99.23 & 99.14 & 99.15 & 99.07 \\ \hline

\end{tabular}

\label{tab:isolation forest}
\end{table}

%Fig. \ref{fig:th_vehicles} depicts the change of throughput against total number of vehicles. Initially, the throughput increases with the increase in a number of vehicles. But when the number of vehicles are close to 20, it becomes to decrease with the increase in number of vehicles. When the number of vehicles are near to 35, the increase or decrease almost get saturated. 
\begin{comment}
 \begin{figure}[h]
\centerline{\includegraphics[scale=0.32]{ThrouhputVsNumberOfVehicles.png}}
\caption{Throughput of the proposed system in terms of  total number of vehicles}
\label{fig:th_vehicles}  
\end{figure}   
\end{comment}

\begin{figure}[h]
\centerline{\includegraphics[scale=0.32]{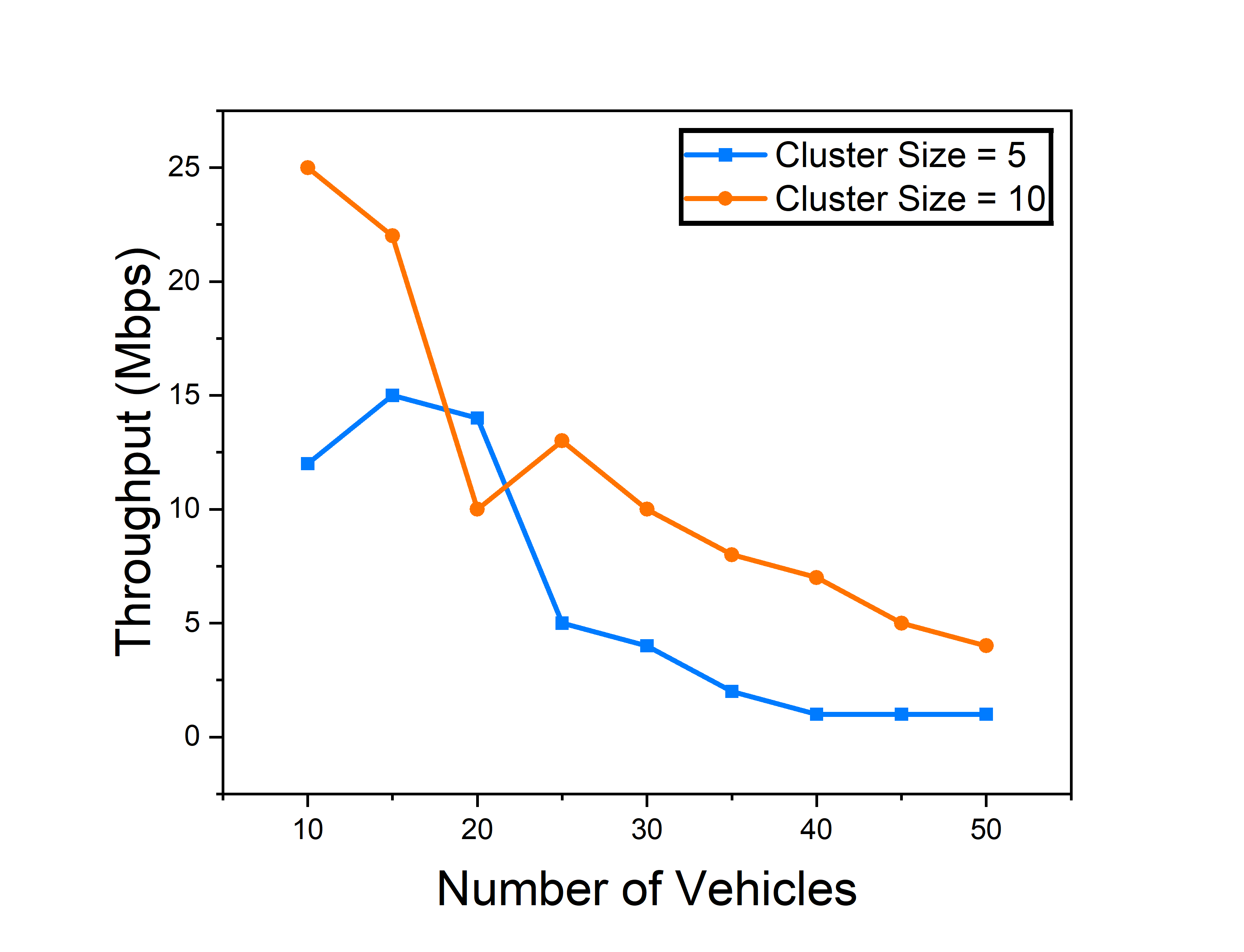}}
\caption{Throughput of the proposed system in terms of different cluster size}
\label{fig:th_cluster}  
\end{figure}

Fig. \ref{fig:th_cluster} illustrates the change in throughput due to the size of the cluster size. Here, we consider two cluster sizes: size of 5 and size of 10. The throughput increases as we increase the cluster size of the proposed system. The gap between the throughput of clusters 5 and 10 is quite big. On the other hand for each cluster size, the throughput decreases for the increase in the number of vehicle nodes. For cluster size 5 the highest throughput was for 15 number of vehicles and it decreases with the increase number of vehicle and reach the lowest point for 50 number of vehicles. On the other hand, the throughput decreases for the limit of 25 vehicles for cluster size 10 and after that it becomes saturated.

\begin{figure}[h]
\centerline{\includegraphics[scale=0.34]{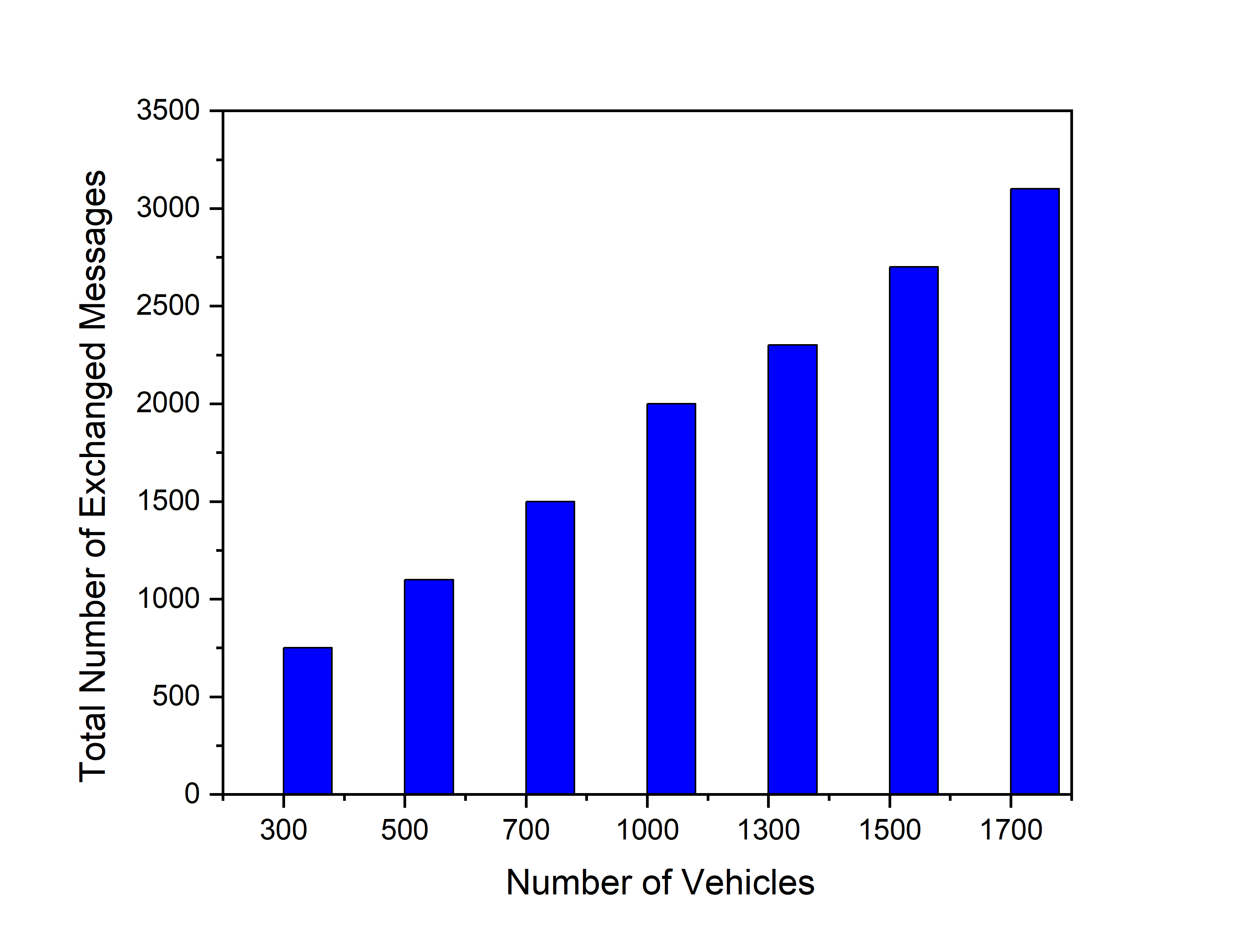}}
\caption{Comparison of communication overhead}
\label{fig:th_overhead}  
\end{figure}

Fig. \ref{fig:th_overhead} depicts the communication cost or the total number of exchanged messages for the proposed system against the total number of vehicles. Communication costs have always increased due to the rise in the number of vehicles.

Table \ref{tab:gas} exhibits the gas consumption with respect of the number of transactions. Gas limit represents the maximum possible required energy for a single transaction. For a small amount of transactions, the gas limit is very low, but it increases linearly for the increasing amount of transactions.

\begin{table}[h!]
\centering
\scriptsize
\caption{Gas consumption against no. of transactions}
\label{tab:gas}
\begin{tabular}{|c|c|}
\hline
\textbf{No. of transactions} & \textbf{Gas Consumption (ETH)} \\ \hline
5  & 27000  \\ \hline
8  & 36000  \\ \hline
10  & 41000  \\ \hline
14 & 55000  \\ \hline
17 & 66000  \\ \hline
22 & 78000  \\ \hline
26 & 89000  \\ \hline
30 & 100000  \\ \hline
\end{tabular}
\end{table}

% \begin{table*}[htb]
%     \centering
%     \scriptsize
%     \caption{Comparative analysis of IEAOCGO-C \cite{elhoseny2023intelligent} technique with proposed Model}
%     \label{tab:net_performance}
%     \begin{tabular}{lccccc}
%         \toprule
%         Vehicles & THRPT(kbps) & PDR(\%) & NLT & ETED & ECM \\
%         \midrule
        
%         -- & Proposed - IEAOCGO‑C  & Proposed - IEAOCGO‑C  & Proposed - IEAOCGO‑C & Proposed - IEAOCGO‑C &  Proposed - IEAOCGO‑C\\
%         \hline
%         20 & 4900 & 31.55 &71.23  & 99.10 &  7.08\\
    
%         \hline
%         30 & 4600 & 52.37 &76.67  & 90.36 & 7.46 \\
%         \hline
 
%         60 & 4200 & 72.00 & 84.56  &84.84  & 7.88 \\
%         \hline
%         80 & 4100 & 85.23 & 88.13 & 75.06 & 8.33 \\
%         \hline
%         100 & 4000  & 91.24  &  90.34& 73.23  & 8.49 \\
%         \bottomrule
%     \end{tabular}
% \end{table*}
Table \ref{tab:net_comparison} shows the comparison of Throughput (THRPT), Packet Delivery Ratio (PDR), Network Lifetime (NLT), End-to-End Delay (ETED), and Energy Consumption (ECM) between our proposed model and IEAOCGO-C algorithm on different vehicle numbers. Our proposed model outperform the IEAOCGO-C \cite{elhoseny2023intelligent}  algorithm on every factor for different vehicle numbers.

\section{Conclusion}

In this research, we have proposed \enquote{DistB-VNET}, a novel framework integrating ml based attack detection, Blockchain, SDN, and NFV to ensure the scalability, security, and efficiency of vehicular communication in smart cities. First, our model filters traffic and blocks malicious ones. Further, our framework utilizes SDN to separate control and data planes while ensuring dynamic resource allocation by NFV and data integrity through an immutable ledger by Blockchain. The results show improved throughput and communication efficiency across different vehicle volumes and cluster sizes. However, the system has not yet been tested in terms of increased transmission costs and gas fees as vehicle numbers and transactions grow. In future, we will focus on optimizing resource management and reducing blockchain transaction costs to improve system scalability.

%------------------------------------------------------------------------------------------------>>>>>
%\section*{List of Abbreviations}%\noindent 
%\begin{tabular}{@{}ll}
%NFV & Network Function Virtualization\\
%SDN & Software Defined Networking\\
%NV & Network Virtualization\\
%VNF & Virtualized Network Functions\\
%OSS & Operation Support Systems\\
%BSS & Business Support Systems\\
%IoT & Internet of Things\\
%%TPP & Trusted Third Party\\
%\end{tabular}

\ifCLASSOPTIONcaptionsoff
  \newpage
\fi

\bibliographystyle{IEEEtran}

\bibliography{sample}

\end{document}